\documentclass[%
 reprint,
 amsmath,amssymb,
 aps,
pra,
]{revtex4-2}

\usepackage{graphicx}
\usepackage{dcolumn}
\usepackage{bm}

\begin{document}


\title{Deep Learning and Density Functional Theory}

\author{Kevin Ryczko}
\email[]{kryczko@uottawa.ca}
\affiliation{Department of Physics, University of Ottawa}

\author{David A. Strubbe}
\affiliation{Department of Physics, University of California, Merced}

\author{Isaac Tamblyn}
\email[]{isaac.tamblyn@nrc.ca}

\affiliation{National Research Council of Canada}
\affiliation{Department of Physics, University of Ottawa}

\date{\today}

\begin{abstract}

We show that deep neural networks can be integrated into, or fully replace, the Kohn-Sham density functional theory scheme for multi-electron systems in simple harmonic oscillator and random external potentials with no feature engineering. We first show that self-consistent charge densities calculated with different exchange-correlation functionals can be used as input to an extensive deep neural network to make predictions for correlation, exchange, external, kinetic and total energies simultaneously. Additionally, we show that one can also make all of the same predictions with the external potential rather than the self-consistent charge density, which allows one to circumvent the Kohn-Sham scheme altogether. We then show that a self-consistent charge density found from a non-local exchange-correlation functional can be used to make energy predictions for a semi-local exchange-correlation functional. Lastly, we use a deep convolutional inverse graphics network to predict the charge density given an external potential for different exchange-correlation functionals and assess the viability of the predicted charge densities. This work shows that extensive deep neural networks are generalizable and transferable given the variability of the potentials (maximum total energy range $\approx100$ Ha), because they require no feature engineering, and because they can scale to an arbitrary system size with an $\mathcal{O}(N)$ computational cost.

\end{abstract}

\maketitle


\begin{figure*}[ht]
	\centering
\includegraphics[width=\textwidth]{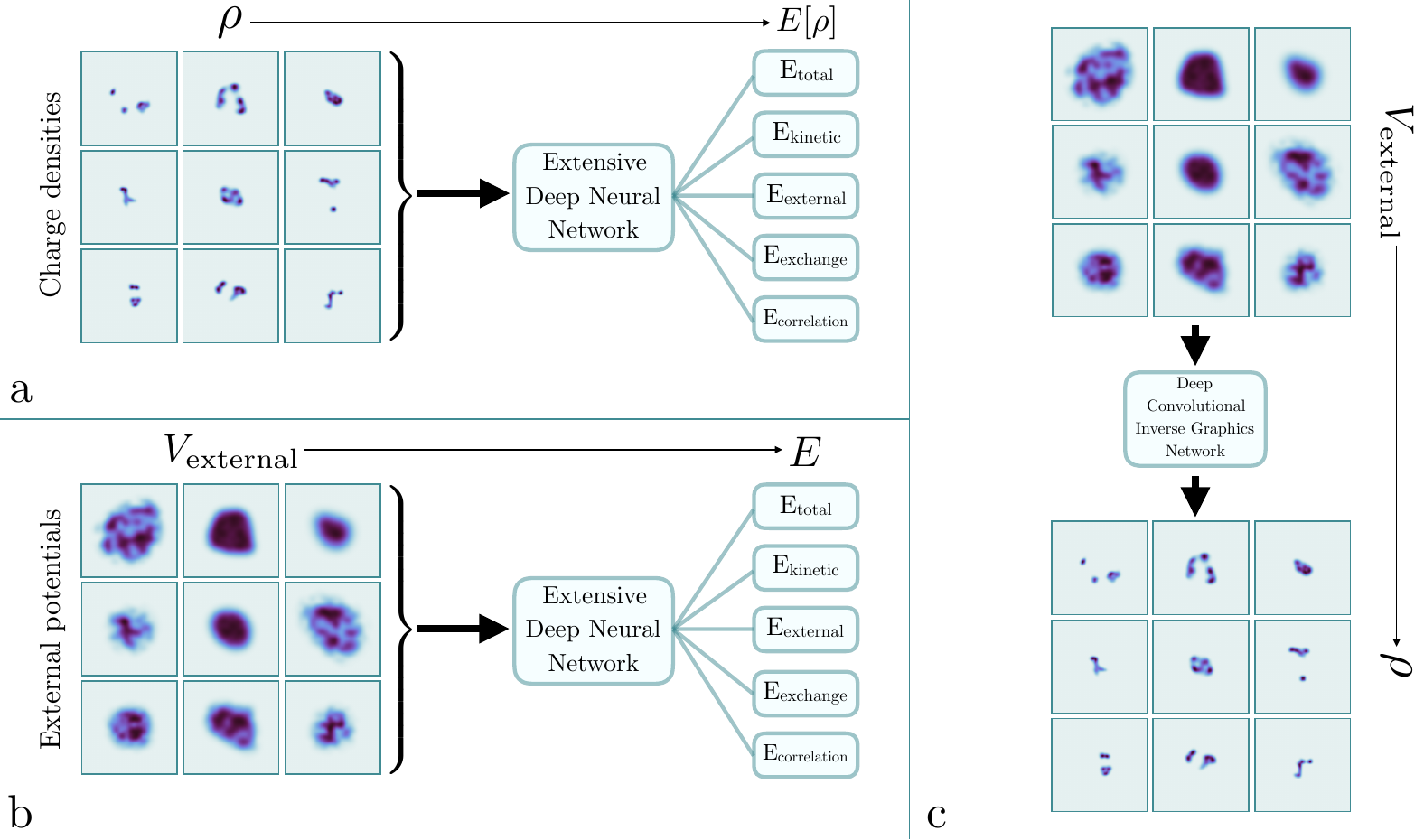}
\caption{\label{networks} A graphical representation that outlines the objectives of this report. In a. and b. we show that both charge densities and external potentials can be used as input to extensive deep neural networks (EDNNs \cite{luchak2017extensive}) to predict the total, kinetic, external, exchange and correlation energies. The images shown are some of the random (RND) potentials along with the self-consistent charge density for that potential. In c. we show that deep convolutional inverse graphics networks (DCIGNs \cite{NIPS2015_5851}) can be used to map external potentials to their respective self-consistent charge densities.}
\end{figure*}

\section{Introduction}
Kohn-Sham (KS) density functional theory (DFT)\cite{kohn1965self} is the standard theoretical tool to study nanoscale systems. Despite its success, DFT calculations for atomistic systems containing tens of thousands to millions of atoms are exceptionally demanding from a computational perspective and are rare in the literature. Machine learning techniques can replace conventional DFT calculations to overcome this computational barrier. Machine learning models are ideal because they rival the accuracy of the method they are trained on, but can be less demanding to evaluate from a computational standpoint. There have been many reports where artificial neural networks (ANNs) have been used to represent potential energy surfaces to accelerate electronic structure calculations \cite{C1CP21668F, LORENZ2004210, doi:10.1021/jp9105585, doi:10.1063/1.3206326, PhysRevLett.98.146401,doi:10.1063/1.5054310, schutt2017schnet, schutt2017quantum}. These reports focus on feature engineering or defining some abstract representation of atomistic systems allowing one to use an ANN. Instead, we focus our review on reports that avoid feature engineering and utilize the electron density in conjunction with machine learning. More specifically, machine learning has become a popular choice to represent energy functionals in DFT \cite{snyder2012finding, PhysRevB.94.245129, LSPH14, SRHB13, osti_1388993}, or to completely circumvent the KS scheme \cite{BVLT17, ryczko2017convolutional}. In deep learning, the machine learning model learns the hierarchical features during training rather than inputing abstract representations. Due to the large number of tuneable parameters in deep neural networks (DNNs) that may include a variety of layers (i.e. convolutional, fully connected, max pooling, etc.), there must be thousands (if not hundreds of thousands) of training examples to find a stable minima with an acceptable accuracy. Generating these training examples is a computationally expensive task, but a trained DNN can evaluate a given quantity at a fraction of a cost compared to the original method.

An alternative, novel approach that has been taken recently by Brockherde \emph{et al.} \cite{BVLT17} is to focus more on uniformly sampling the space that a machine learning model will eventually predict and to use traditional machine learning with far fewer tuneable parameters. This approach was successful in predicting KS-DFT total energies and charge densities in one dimension (1D) for random Gaussian potentials and for small molecules in three dimensions (3D). Due to the use of Kernel Ridge Regression (KRR), they were able to achieve an acceptable accuracy with a relatively small number of training examples. Unfortunately, KRR is known to have poor scaling with respect to the number of training examples, making it difficult to train with a large (and more diverse) set of training examples.

In KS-DFT, one of the contributions to the total energy is the non-interacting kinetic energy. Before the KS scheme was realized, Hohenberg and Kohn \cite{hohenberg1964inhomogeneous} postulated the formalism for an interacting kinetic energy functional of the density. An analytic expression for the exact interacting kinetic energy functional with respect to the electron density is unknown. This is one of the major downfalls of orbital-free (OF) DFT, where all energy contributions are explicitly written in terms of the electron density. This shortcoming provides motivation to construct an approximate functional of the density with a machine learning model. In a report done by Yao \emph{et al.} \cite{yao2016kinetic}, a convolutional neural network (CNN) was used to represent the kinetic energy functional in the OF-DFT total energy expression for various hydrocarbons. Their data generation process consisted of performing KS-DFT and collecting the charge density along with the KS non-interacting kinetic energy. The charge density was then used as input to the CNN with the KS non-interacting kinetic energy as the label. With this representation they were able to successfully reproduce potential energy surfaces when compared to the true KS potential energy surfaces. In another report by Snyder \emph{et al.} \cite{SRHB13}, they were able to use a machine learning model to make kinetic energy predictions given a charge density for a diatomic molecule. Using their framework, they were able to accurately dissociate the diatomic molecule, and compute forces suggesting that \emph{ab initio} molecular dynamics could eventually be done via machine learning methods.

When representing the kinetic energy with a machine learning model in the OF scheme, one then becomes concerned with calculating the functional derivative of the machine learning model with respect to the density. In a report from Li \emph{et al.} \cite{LSPH14}, they showed there is a trade-off between accuracy and numerical noise to when taking the functional derivative of a machine learning model. Brockherde \emph{et al.} \cite{BVLT17} avoided this issue by training a machine learning model to learn the mapping between the potential and the electron density, avoiding the functional derivative.

In another recent report by Kolb \emph{et al.} \cite{osti_1388993}, a software package was developed to combine artificial neural networks with electronic structure calculations and molecular dynamics engines. Using their newly developed software, they were able to show that artificial neural networks can be used to make predictions with the electronic charge density as input and various energies as output. Specifically, they were able to predict energies and band gaps calculated at a higher level of theory from charge densities calculated at a lower level of theory. This approach is very advantageous as high level theory calculations (i.e. $G_0W_0$ \cite{PhysRevB.34.5390}) become quite computationally expensive for larger systems.

Although significant progress has been made incorporating machine learning and deep learning to a variety of electronic structure problems, most do not have the ability to properly handle extensive properties. In some of our past work, we introduced extensive DNNs (EDNNs) \cite{luchak2017extensive} to intrinsically learn extensive properties. This means that when the DNN learns the fundamental screening length scale it can then easily scale up to massive systems in a trivially parallel manner. EDNNs work by first dividing up an image into fragments which are called focus regions. These fragments are then padded with context regions. The context regions may also respect periodic boundary conditions. Each of these fragments can then be simultaneously passed into machine learning models that share weights. It should be noted here that any machine learning method that uses back propagation to minimize the loss function can be used. Finally, the outputs of the machine learning models are then summed yielding the final prediction from the EDNN.

In this report, we show that EDNNs have the capability to learn energy and charge density mappings that could replace some, if not all, calculations in KS-DFT scheme. We push the frontier of what EDNNs can learn from charge densities and external potentials by calculating the self-consistent charge densities in external potentials with extreme variabilities. In previous reports that focus on small molecules \cite{BVLT17, doi:10.1063/1.3206326, osti_1388993, yao2016kinetic}, the self-consistent charge densities generated from molecular dynamics are similar and have small energy ranges (i.e. $\approx31.8$ mHa \cite{BVLT17}). We avoid small molecules (where the charge density would be localized in space), and truly challenge the ability of EDNNs to make accurate predictions across a variety of electronic environments. Quantitatively speaking the energy range of our 10 electron calculations with our random (RND) external potentials is $\approx 100$ Ha. This report is outlined as follows: In Section \ref{methods}, we describe our data generation process, as well as the DNN topologies and hyper-parameter selections. In Subsection \ref{energy_predictions}, we show that DNNs have the capability to act as density functionals and can accurately predict the exchange, correlation, external, kinetic, and total energies \emph{simultaneously} (Subsection \ref{functional}). We also show that EDNNs can also circumvent the KS scheme (Subsection \ref{by_pass}) by mapping the external potential to all of the aforementioned energies simultaneously. Additionally, we show that EDNNs can be used in a somewhat ``perturbative'' manner, where we predict energies computed with semi-local or non-local exchange-correlation functionals from non-local electron densities. In Subsection \ref{image_predictions}, we show that deep convolutional inverse graphics networks (DCIGNs) can also map the external potential to the electron density, and assess the viability of the predicted electron density. 
Lastly, in Section \ref{conclusion}, we summarize our results and consider future work that could be done with our new framework. The outline of this manuscript can be seen graphically in Figure \ref{networks}.

\begin{figure*}[ht]
	\centering
\includegraphics[width=\textwidth]{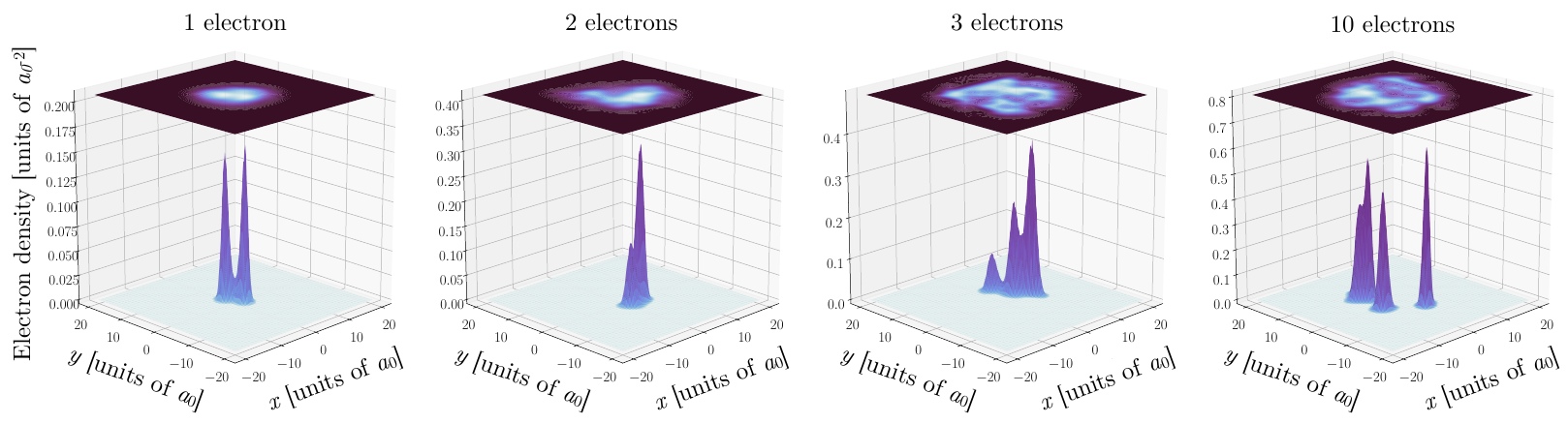}
\caption{\label{HO_cd_pot} Computed charge densities (3D surfaces) and random external potential energy surfaces (2D surfaces) for typical configurations of systems with 1, 2, 3, and 10 electrons.}
\end{figure*}

\section{Methods}
\label{methods}
We investigate two-dimensional (2D) electron gases within the KS-DFT framework \cite{kohn1965self} for two external potentials: simple harmonic oscillator (SHO) and RND. These external potentials have been used in a previous study \cite{PhysRevA.96.042113} for direct diagonalization, one-electron calculations.
 In the KS-DFT framework, one minimizes the total energy functional
\begin{equation}
	\label{KS_func}
	E[\rho]=T[\rho] + E_{\text{ext}}[\rho] + E_{\text{Hartree}}[\rho] + E_{\text{XC}}[\rho].
\end{equation} 
In Equation \ref{KS_func}, $T$ is the non-interacting kinetic energy, $E_{\text{ext}}$ is the energy due to the interaction of the electrons with the external potential, $E_{\text{Hartree}}$ is the electrostatic energy describing the electron-electron interactions, and $E_{\text{XC}}$ is the exchange-correlation energy,

Using EDNNs, we investigate the feasibility of learning the total energy as well as the individual contributions to the total energy. We therefore have trained models to predict the total, non-interacting kinetic, external, exchange, and correlation energies. The external potentials chosen for this report, as mentioned previously, are SHO and RND potentials in 2D. The SHO potentials take the form 
\begin{equation}
	V_{\text{ext}}(\{x_i\})=\frac{1}{2}\sum_i^{D}k_i(x_i - x_{0_i})^2
\end{equation}
where $D$ is the dimension, $k_i=m\omega_i^2$ is the spring constant, and $x_{0_i}$ is the shift of the potential in a given coordinate. For the RND potentials we follow the work of Mills \emph{et al.} \cite{PhysRevA.96.042113} when generating the potentials on a grid.
We refer the reader to the original manuscript \cite{PhysRevA.96.042113} for more information on the RND potential generation. To briefly summarize the process, the first step consists of generating a random binary image of 0's and 1's and then applying a gaussian filter. The second step consists of generating a mask that is constructed with a random convex hull and an additional gaussian blur. The mask is then applied onto the image yielding the final result. The larger energy scale of the RND external potentials can be attributed to the length scales of the RND external potentials. The average Gaussian kernel sizes in the external potential generation is $\sim3$ Bohr, whereas the average length scale of the SHO external potentials is $\sim12$ Bohr. Assuming that the energy scales as $E\sim1/r^2$, the energy scale of the RND external potentials is 16 times larger than the SHO external potential energy scale on average. To create datasets large enough to use DNNs, we chose to randomly sample $k_i$ and $x_{0_i}$ such that $0.01\leq k_i\leq 0.16 \text{ Ha} / a_0^2 $ (Hartree per Bohr${}^2$) and $-8.0\leq x_{0_i}\leq 8.0~a_0$. With a given selection of these variables, the external potential was then evaluated on a $40\times40~a_0$ space with a $256\times256$ grid point mesh. 
We then chose to place either $N=$1, 2, 3, or 10 electrons in the 2D space. For each choice of the number of electrons, we generate an external potential, and then perform three DFT calculations, each with different exchange-correlation functionals. We used the local density approximation (LDA) Slater exchange functional in 2D \cite{dirac_1930, bloch1929bemerkung}, the Perdew-Burke-Ernzerhof (PBE) \cite{perdew1996generalized} generalized-gradient approximation (GGA) exchange functional, and  and the 2D meta-generalized gradient approximation (MGGA) exchange functional from Pittalis \emph{et al.} \cite{pittalis2007exchange}, in each case paired with the correlation functional of Attaccalide \emph{et al.} \cite{attaccalite2002correlation}, all provided by the libxc library \cite{marques2012libxc, lehtola2018recent}. Here, we do not take the orbitals or charge densities from the LDA calculations to calculate energies at the PBE or MGGA level. All of the energies for each functional are calculated independently.
 All of the calculations were carried in real space with the Octopus code \cite{andrade2015real, andrade2012time}. For testing, we set aside 10\% of each data set. This made for 90,000 training configurations and 10,000 testing configurations for each case of potential, number of electrons, and exchange-correlation functional. Note that the number of electrons and the type of external potential uniquely defines a dataset. Therefore, the 10\% set aside for testing includes 10,000 external potentials and 30,000 charge densities (10,000 for each exchange-correlation functional choice). In addition, the labels were normalized independently such that each of the components had a range from zero to one. The calculations are summarized in Table \ref{calcs_summary} of Appendix \ref{data_info}. All data used in this report is available online (\url{http://clean.energyscience.ca/datasets}) along with the code (\url{https://github.com/CLEANit/multi-scale-ednn}) to allow for future development of featureless deep learning based functionals.

When constructing the EDNNs, we used a mixture of Tensorflow \cite{abadi2016tensorflow} and TFLearn \cite{tang2016tf} in Python. For the networks topologies we build on our previous reports \cite{ryczko2017convolutional, luchak2017extensive} and use EDNNs where each tile of the EDNN has the same in-tile CNN used previously for predicting KS-DFT total energies of 2D hexagonal sheets \cite{ryczko2017convolutional}. For clarity, the in-tile CNN consisted of 2 reducing convolutional layers with kernel sizes of 3, 6 non-reducing convolutional layers with kernel sizes of 4, 1 reducing convolutional layer with a kernel size of 3, 4 non-reducing convolutional layers with kernel sizes of 3, a fully connected layer with 1024 neurons, and a final fully connected layer with one neuron. All of the activations used were rectified linear units. We emphasize that in our approach, we do not do any sort of feature engineering, like past reports that use ANNs \cite{doi:10.1063/1.5054310, PhysRevLett.98.146401, doi:10.1063/1.3206326, doi:10.1021/jp9105585, LORENZ2004210, C1CP21668F, osti_1388993}. The convolutional layers in the EDNNs identify relevant features during the training process. When utilizing an EDNN, one must declare the focus and context regions which is used to ``tile" up the image into fragments. To find the ideal focus and context regions, we started by training the EDNNs on the 2D charge density to total energy mapping as well as the 2D external potential to total energy mapping for the 1, 2, 3, and 10 electron systems for calculations done with the LDA exchange-correlation functional and the SHO external potential. We chose a variety of focus and context sizes, and found that the optimal focus and context sizes are 128 pixels for the focus size, and 32 pixels for the context size. Our decision was based on a balance between accuracy and computation time. A larger focus size lowers the computation time, and a larger context size yields larger images, resulting in more neurons in the EDNNs thereby improving the accuracy of the model. For a focus of 128 pixels, we found that the accuracies were very similar for various context sizes and the choice of 32 pixels was almost arbitrary. This hyperparameter search was the most computationally demanding task for this work due to the number of models that had to be trained. While training, we used a learning rate of $10^{-4}$ for 500 epochs when using the charge densities as input and a learning rate of $10^{-5}$ for 500 epochs when using the external potentials as input. In both cases, we further reduced the learning rates by a factor of 10 and trained for an additional 100 epochs. For clarity, an epoch is defined  to be when the weights of the network have been updated for the entire training dataset once. 

\begin{figure*}[ht]
	\centering
\includegraphics[width=\textwidth]{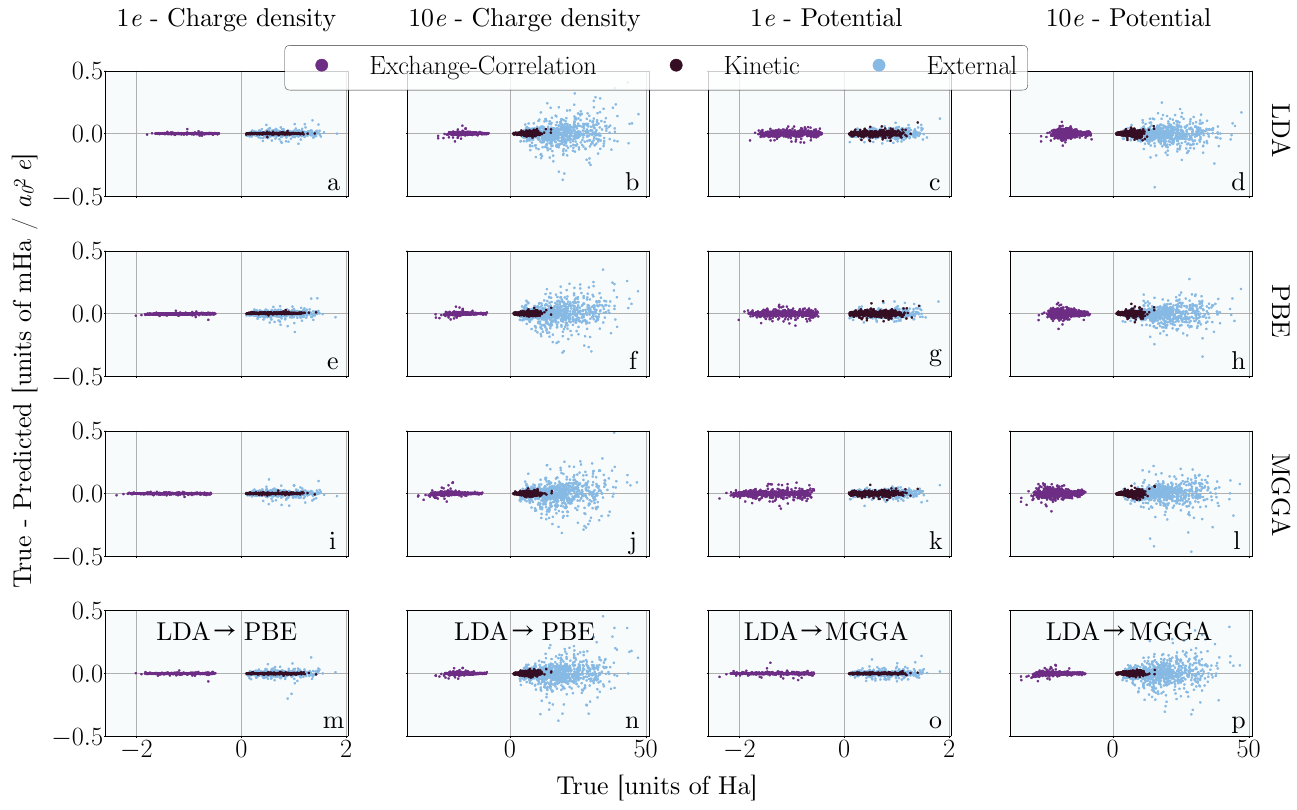}
\caption{\label{main_results_1} True minus predicted (in mHa / $a_0^2$ / electron) versus true (Ha) for various models with the RND potentials. Plots a-d are models trained with the LDA exchange-correlation functional, e-f with the PBE exchange-correlation functional, and i-l with the MGGA exchange-correlation functional. First column (a, e, i) is for 1 electron models where the charge densities were used as input. Second column (b, f, j) is for 10 electron models where the charge densities were used as input. Third column (c, g, k) is for 1 electron models where the external potentials were used as input. Fourth column (d, h, l) is for 10 electrons models where the external potentials were used as input. The bottom row (m-p) is for models where LDA charge densities were used as input, and the labels were either PBE energies (m, n) or MGGA energies (o, p). Plots m, o are for the 1 electron systems, and n, p for the 10 electron systems. It should be noted that one model is predicting the correlation, exchange, external, kinetic, and total energies. We have combined the exchange-correlation error and omitted the total energy error for clarity.}
\end{figure*}

\section{Results}

\label{results}
\subsection{Energy predictions}
\label{energy_predictions}
\subsubsection{EDNNs as a functional}
\label{functional}
Firstly, we show that EDNNs can be used as an energy functional for correlation, exchange, external, kinetic and total energies. For the LDA, PBE, and MGGA functionals discussed in Section \ref{methods}, we used the computed self-consistent charge densities as input to an EDNN and were able to successfully predict the correlation, exchange, external and total energies simultaneously for both SHO and RND external potentials. Starting with the models where the SHO external potentials were used in the DFT calculations, we found that the mean absolute errors for each particular case are less than 1.5 mHa. These can be seen in Table \ref{main_results_2_table} of Appendix \ref{mae_sho}. In Figure \ref{main_results_1}, we show predicted minus true versus true for the one and ten electron models with the different exchange-correlation functionals when the RND external potentials were used in the DFT calculations. In this Figure, it is clear that the error of the models increase with the number of electrons. This increase in error is expected due to the increase in the range of energies and can be physically attributed to the increase of interactions in the system. Looking to Table \ref{main_results_1_table} of Appendix \ref{mae_rnd} we also observe that the mean absolute errors become larger as the complexity of the exchange-correlation functional increases. In addition to these trends, we also notice that the energy with the largest mean absolute error comes from the external energy functional. This again can be attributed to the ranges of the various energies. The external energy has the largest range of all the energies being predicted. We also address the generalizability of the models by testing the model that was trained on 10 electron charge densities calculated with the RND external potentials and the LDA functional with 10 electron charge densities calculated with the SHO external potentials and the LDA functional (and vice versa). We found in both cases the errors increased by several orders of magnitude. This is not surprising given the different energy ranges of the datasets. On the contrary, when examining the true versus predicted plot for the model trained on the RND dataset but tested on with the SHO dataset, we found that a constant shift could simply be added to substantially decrease the error. We expect that this constant shift could be easily rectified if SHO training examples are included in the training process. 

DFT is a more popular choice for larger systems relative to wavefunction based methods because the exchange-correlation functionals used are computationally inexpensive relative to methods that employ exact exchange, for example. In light of this, we have trained EDNNs to predict energies at the PBE and MGGA level given a self-consistent charge density computed with the LDA exchange-correlation functional. In Figure \ref{main_results_1}, we consider 1 and 10 electron models trained on the mapping between LDA charge densities and either PBE or MGGA energies. Similar to the results mentioned above, the mean absolute errors increase both with the number of electrons and the complexity of the exchange-correlation functional. In Table \ref{main_results_1_table} of Appendix \ref{mae_rnd}, we also notice that the highest mean absolute error is for the external and total energies. This result further suggests that there is not a fundamental problem with learning the external energy, but the larger range of energies makes it more difficult for a EDNN to handle with extreme precision. In addition, since the correlation functional is the same across all of the calculations and the same testing data was used for each case of number of electrons, we can determine if the networks are learning the correlation energy in a similar manner. In Table \ref{main_results_1_table}, we can see that the correlation energies have similar magnitudes of error indicating that similar correlation functional mappings are being learned as one should expect. The success of learning the energies of a more accurate exchange-correlation functional given a less accurate charge density shows promise for other applications. A future application could include learning $G_0W_0$ quasiparticle energies from a DFT computed self-consistent charge density, similar to the work that was completed by Kolb \emph{et al.} \cite{osti_1388993}.

A note should be made about Table \ref{main_results_1_table} with respect to the magnitude of some of the mean absolute errors reported. In comparison to the report by Mills \emph{et al.} \cite{PhysRevA.96.042113}, some of the mean absolute errors are larger by some cases a factor of 10. In addition, the focus and context hyperparameters were optimized for the SHO external potentials. In the work of Brockherde \emph{et al.} \cite{BVLT17}, they managed to reach chemical accuracy using three dimensional charge densities, but the energy range of their training set was $\approx$40 kcal/mol (for a benzene molecule). For their best reported model with a mean absolute error of 0.28 kcal/mol, the relative mean absolute error, that we define as the mean absolute error divided by the range of the dataset is 0.007. In our 10 electron model with the RND external potential, our energy range was $\approx$100 Ha (62750.9 kcal/mol) and the mean absolute error of the total energy predictions was 78.514 mHa yielding a relative mean absolute error of 7.85$\times10^{-4}$. 
 

\subsubsection{Circumventing Kohn-Sham DFT}
\label{by_pass}
In addition to using EDNNs as a functional, it is arguably more convenient to train a EDNN to learn the mapping between the external potential and the contributing energies of that system. It is more convenient because it avoids calculating a self-consistent charge density with the KS scheme. We have trained EDNNs to predict the exchange, correlation, external, kinetic, and total energy simultaneously using the external potential as input rather than the charge density. Again, in Figure \ref{main_results_1} we show true minus predicted versus true for the correlation, exchange, external, kinetic, total energies for the RND external potentials. Here, it is evident that the charge density is more optimal as an input to an EDNN for the 1 electron systems. There is much more spread in the distribution when using external potentials as input compared to charge densities. For 10 electrons, this is not the case. Looking to Table \ref{main_results_1_table} of Appendix \ref{mae_rnd}, we can see that for 1, 2, and 3 electrons no matter what choice of exchange-correlation functional, it is less difficult to learn the mapping between $\rho\rightarrow E$ than $V\rightarrow E$. The mean absolute errors are lower for all energies. In the case of 10 electrons, the mean absolute errors in the external and total energies are lower for the models that have potentials as input. Although the errors are lower for the external and total energies, the mean absolute errors for correlation, exchange, and kinetic energies are larger. When training a model on a set of energies, there is a balance between the errors of the energies since the loss function depends on the sum over the mean squared errors between the true and predicted energies. In the case of using charge densities as input to the EDNN, we found the exchange, correlation, and kinetic energies can be predicted with much better accuracy than the external or total energies. In the case of using potentials as input to the EDNN, we found that there is more of a balance of accuracy between the different energies being predicted.


\subsection{Image predictions with Deep Convolutional Inverse Graphics Networks}
\label{image_predictions}
\begin{figure}[ht]
	\centering
\includegraphics[width=0.8\linewidth]{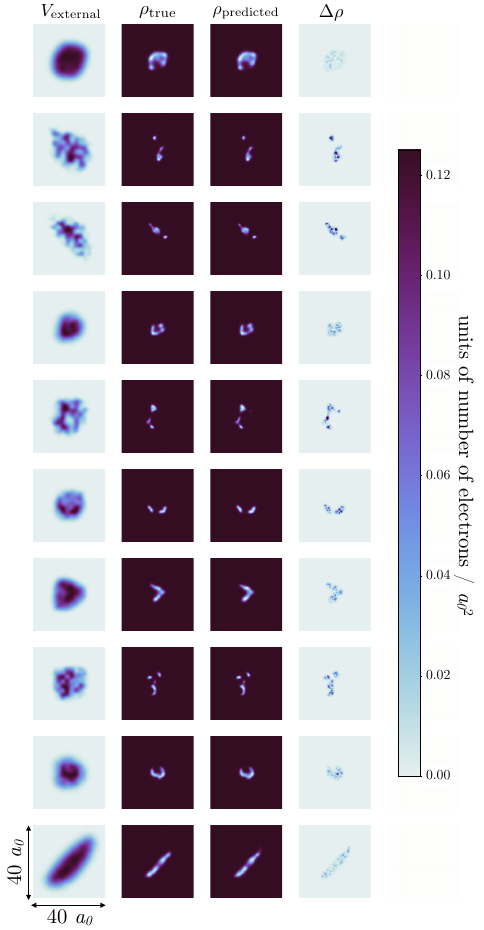}
\caption{\label{predicted_wfs} Examples of the RND potentials, true charge densities $(\rho_{\text{true}})$, predicted charge densities $(\rho_{\text{predicted}})$, and differences between the true and predicted charge densities $(\Delta\rho)$. The charge densities shown here were computed with the LDA exchange-correlation functional. The colour bar is for the charge density differences $\Delta\rho$.}
\end{figure}

In both KS-DFT and OF-DFT, the self-consistent charge density is the central quantity that one is interested in calculating. Once one has the charge density, most other quantities can be calculated in a straightforward manner. In this Subsection, we address the viability of using DCIGNs to map the external potential to the self-consistent charge density in 2D for the RND potentials with the LDA, PBE, and MGGA exchange correlation functionals. DCIGNs were recently introduced in the literature \cite{NIPS2015_5851} and have a similar topology to autoencoders \cite{Goodfellow-et-al-2016}. The DCIGN that we have used has 4 reducing convolutional layers, 3 non-reducing convolutional layers, and 4 deconvolutional layers such that the output image has the same dimensionality as the input image. This topology differs slightly from the original work on DCIGNs \cite{NIPS2015_5851}, where a fully connected would replace our 3 non-reducing convolutional layers. Additionally, our DCIGN is deterministic. In the original work \cite{NIPS2015_5851}, random noise is introduced in the decoder to create a non-deterministic generative model. All of our convolutional layers use a kernel size of 3 with rectified linear unit activations. We used a learning rate of $10^{-5}$ while training for 500 epochs and dropped the learning rate by a factor of 10 before training for an additional 100 epochs. For this discussion we focus solely on the 10 electron calculations with the RND external potentials. We argue that these are the most challenging calculations to train with a DCIGN, and can therefore safely assume that the less complex calculations would be successful given the success of the most complex cases. In Figure \ref{predicted_wfs}, we show some of the predictions that the DCIGN made for 10 electrons calculations with the LDA exchange-correlation functional. There is a remarkable resemblance between the true ($\rho_{\text{true}}$) and predicted ($\rho_{\text{predicted}}$) charge densities. The DCIGN is capable of handling the extreme variability of the complex shapes, and is capable of handling the cases where the charge density is not isolated to one region of space. From a qualitative perspective, the DCIGN makes accurate predictions of the charge densities given RND external potentials. 

Normally, when addressing the viability of a machine learning model from a quantitative perspective one considers the mean absolute error on the test set. We argue that a more rigorous test for $\rho_{\text{predicted}}$ would be mean absolute error of the energies associated with $\rho_{\text{predicted}}$. We therefore take $\rho_{\text{predicted}}$ and renormalize them such that $\int d{\bf r}~ \rho({\bf r})_{\text{predicted}}=10$. Afterwards, we use the renormalized $\rho_{\text{predicted}}$ as input to a subset of the models described in Subsection \ref{functional}. We then compare the energies predicted from $\rho_{\text{predicted}}$ with the true energies. In Table \ref{main_results_1_table} of Appendix \ref{mae_rnd}, we show the mean absolute errors between the true and predicted energies for the different exchange-correlation functionals. When comparing the mean absolute errors of the predicted energies for $\rho_{\text{predicted}}$ with the energy predictions made from the $\rho_{\text{true}}$, the minimal difference was seen for the correlation energies with a value of $\sim 6$ mHa. This was true for all exchange-correlation functionals considered in this work. The maximal difference between the mean absolute errors when comparing the energy predictions of $\rho_{\text{predicted}}$ and $\rho_{\text{true}}$ was the total energy which was $\sim 20$ mHa. Again, this is true for all exchange-correlation functionals considered. In addition to this metric, we also report the mean absolute error for a model mapping $\rho_{\text{predicted}}$ to the true energies. We find an increase in the errors in comparison to the models that map the true charge densities to the true energies, but the errors are comparable to the errors when we evaluate $\rho_{\text{predicted}}$ with the model trained on the true charge densities. Training with the charge densities and energies as labels yields similar results. We also report the mean absolute density driven error (DDE) using a similar definition as Brockherde \emph{et al.} \cite{BVLT17} in Table \ref{main_results_1_table}. We find similar trends in the DDEs when comparing them to the mean absolute errors of $\rho_{\text{predicted}}$. For more information on our definition of the mean absolute DDEs, see Appendix \ref{MADDE}. In addition, to compare with Brockherde \emph{et al.} \cite{BVLT17}, our relative mean absolute error is a factor of $\approx7$ times smaller.

\section{Conclusion}
\label{conclusion}
In conclusion, we have shown that EDNNs and DCIGNs can be used alongside, or replace conventional KS-DFT calculations. For both the RND and SHO external potentials, EDNNs have the capability to make highly accurate energy predictions using both the charge densities and the external potentials as input for correlation, exchange, external, kinetic and total energy simultaneously (dataset is available online here: \url{http://clean.energyscience.ca/datasets}). In addition, we have shown that DCIGNs have the capability to predict charge densities given an external potential. Qualitatively speaking, the predicted charge densities are remarkably similar to the true charge densities. Quantitatively speaking, the relative mean absolute errors were found to be smaller than previous, state-of-the-art work \cite{BVLT17}. The results of this report show promise for future application in two regards. First, that this framework has the capability to make predictions of higher level theory calculations given a lower level theory charge density similar to a previous report \cite{osti_1388993}. Second, both EDNNs and DCIGNs can be used to calculate energies covering a large range of electronic environments to a high level of accuracy.

\section{Acknowledgements}
K. R. and I. T. acknowledges funding from the Natural Sciences and Engineering Research Council of Canada, and Compute Canada and SOSCIP for computational resources. K. R. and I. T. would also like to thank NVIDIA for a faculty hardware grant. D. A. S. acknowledges University of California, Merced start-up funding.
\bibliography{refs}

\appendix

\section{More details on the generated data}
\label{data_info}
\begin{center}
	\begin{table}[ht]
 	\begin{tabular}{c|c|c|c}
		$N$ & $V_{\text{ext}}$ & $V_{\text{X}} + V_{\text{C}}$ & Number of calculations \\ \hline
		1 & SHO & LDA + LDA & 100,000 \\
		1 & SHO & PBE + LDA & 100,000 \\
		1 & SHO & MGGA + LDA & 100,000 \\
		1 & RND & LDA + LDA & 100,000 \\
		1 & RND & PBE + LDA & 100,000 \\
		1 & RND & MGGA + LDA & 100,000 \\
		2 & SHO & LDA + LDA & 100,000 \\
		2 & SHO & PBE + LDA & 100,000 \\
		2 & SHO & MGGA + LDA & 100,000 \\
		2 & RND & LDA + LDA & 100,000 \\
		2 & RND & PBE + LDA & 100,000 \\
		2 & RND & MGGA + LDA & 100,000 \\
		3 & SHO & LDA + LDA & 100,000 \\
		3 & SHO & PBE + LDA & 100,000 \\
		3 & SHO & MGGA + LDA & 100,000 \\
		3 & RND & LDA + LDA & 100,000 \\
		3 & RND & PBE + LDA & 100,000 \\
		3 & RND & MGGA + LDA & 100,000 \\
		10 & SHO & LDA + LDA & 100,000 \\
		10 & SHO & PBE + LDA & 100,000 \\
		10 & SHO & MGGA + LDA & 100,000 \\
		10 & RND & LDA + LDA & 100,000 \\
		10 & RND & PBE + LDA & 100,000 \\
		10 & RND & MGGA + LDA & 100,000 \\ \hline
		total & & & 2,400,000
	\end{tabular}
	\caption{\label{calcs_summary} Summary of the calculations that were used for training and testing the deep learning models. $N$ is the number of electrons, $V_{\text{ext}}$ is the external potential chosen (see text), and $V_{\text{X}} + V_{\text{C}}$ are the exchange-correlation potentials chosen. Note that the combination of the number of electrons and external potential produces a unique data set. For example, the 3 electron systems in RND potentials has 100,000 external potentials but contributes 300,000 calculations due to the use of different exchange-correlation functionals. }
	\end{table}
\end{center}

\clearpage

\section{Mean absolute errors with the simple harmonic oscillator external potentials}
\label{mae_sho}
\begin{table*}[ht]
	\begin{tabular}{c|c|c|c|c|c|c|c}
	$N_{\text{electrons}}$ & Input & Functional & $E_{\text{correlation}}$  &  $E_{\text{exchange}}$  & $E_{\text{external}}$  & $E_{\text{kinetic}}$  & $E_{\text{total}}$   \\ \hline
		1 & $\rho$  & LDA  & 0.1 (0.1) & 0.1 (0.1) & 0.1 (0.1) & 0.1 (0.1) & 0.2 (0.2)   \\
		2 & $\rho$  & LDA  & 0.1 (0.1) & 0.1 (0.2) & 0.1 (0.2) & 0.1 (0.1) & 0.3 (0.4) \\
		3 & $\rho$  & LDA  & 0.1 (0.1) & 0.1 (0.2) & 0.2 (0.3) & 0.1 (0.2) & 0.5 (0.6) \\
	 10 & $\rho$  & LDA   & 0.1 (0.1) & 0.2 (0.2) & 0.3 (0.6) & 0.2 (0.3) & 0.9 (1.4) \\ \hline
		1 & $\rho$  & PBE  & 0.1 (0.1) & 0.2 (0.2) & 0.1 (0.2) & 0.1 (0.1) & 0.2 (0.3) \\
		2 & $\rho$  & PBE  & 0.1 (0.1) & 0.2 (0.2) & 0.1 (0.2) & 0.1 (0.1) & 0.3 (0.4) \\
		3 & $\rho$  & PBE  & 0.1 (0.2) & 0.2 (0.3) & 0.3 (0.4) & 0.2 (0.2) & 0.8 (0.9) \\
	   10 & $\rho$  & PBE  & 0.1 (0.1) & 0.2 (0.2) & 0.4 (0.6) & 0.2 (0.2) & 0.9 (1.5) \\ \hline
		1 & $\rho$  & MGGA   & 0.1 (0.1) & 0.2 (0.3) & 0.1 (0.2) & 0.1 (0.1) & 0.3 (0.3) \\
		2 & $\rho$  & MGGA   & 0.2 (0.2) & 0.3 (0.4) & 0.2 (0.2) & 0.1 (0.2) & 0.4 (0.5) \\
		3 & $\rho$  & MGGA    & 0.1 (0.1) & 0.2 (0.3) & 0.2 (0.2) & 0.1 (0.2) & 0.4 (0.5) \\
	   10 & $\rho$  & MGGA   & 0.2 (0.3) & 0.4 (0.5) & 0.5 (0.8) & 0.3 (0.4) & 1.3 (1.9)\\ \hline
		1 & $V_{\text{ext}}$ & LDA   & 0.1 (0.1) & 0.1 (0.1) & 0.1 (0.1) & 0.1 (0.1) & 0.2 (0.2) \\
		2 & $V_{\text{ext}}$ & LDA   & 0.1 (0.1) & 0.1 (0.2) & 0.1 (0.1) & 0.1 (0.1) & 0.2 (0.3) \\
		3 & $V_{\text{ext}}$ & LDA   & 0.1 (0.1) & 0.1 (0.2) & 0.1 (0.2) & 0.1 (0.1) & 0.3 (0.4) \\
	   10 & $V_{\text{ext}}$ & LDA  & 0.1 (0.2) & 0.2 (0.3) & 0.3 (0.8) & 0.2 (0.3) & 0.8 (1.1) \\ \hline
		1 & $V_{\text{ext}}$ & PBE   & 0.1 (0.1) & 0.1 (0.1) & 0.1 (0.1) & 0.1 (0.1) & 0.1 (0.2) \\
		2 & $V_{\text{ext}}$ & PBE   & 0.1 (0.1) & 0.1 (0.2) & 0.1 (0.1) & 0.0 (0.1) & 0.2 (0.3) \\
		3 & $V_{\text{ext}}$ & PBE   & 0.1 (0.1) & 0.1 (0.2) & 0.1 (0.2) & 0.1 (0.1) & 0.3 (0.4) \\
	   10 & $V_{\text{ext}}$ & PBE  & 0.1 (0.2) & 0.2 (0.4) & 0.4 (0.8) & 0.2 (0.3) & 0.8 (1.1) \\ \hline
		1 & $V_{\text{ext}}$ & MGGA  & 0.1 (0.1) & 0.1 (0.2) & 0.1 (0.1) & 0.1 (0.1) & 0.2 (0.2) \\
		2 & $V_{\text{ext}}$ & MGGA  & 0.1 (0.1) & 0.2 (0.2) & 0.1 (0.1) & 0.0 (0.1) & 0.2 (0.3) \\
		3 & $V_{\text{ext}}$ & MGGA  & 0.1 (0.1) & 0.2 (0.3) & 0.1 (0.2) & 0.1 (0.1) & 0.3 (0.4) \\
	   10 & $V_{\text{ext}}$ & MGGA  & 0.1 (0.2) & 0.3 (0.5) & 0.3 (0.7) & 0.2 (0.3) & 0.7 (1.0) \\ \hline
 		1 & $\rho$  & LDA$\rightarrow$PBE  & 0.1 (0.1) & 0.1 (0.2) & 0.1 (0.1) & 0.1 (0.1) & 0.2 (0.3)\\
 		2 & $\rho$  & LDA$\rightarrow$PBE  & 0.1 (0.1) & 0.2 (0.2) & 0.1 (0.2) & 0.1 (0.1) & 0.3 (0.4) \\
 		3 & $\rho$  & LDA$\rightarrow$PBE   & 0.1 (0.1) & 0.1 (0.2) & 0.2 (0.2) & 0.1 (0.2) & 0.4 (0.5) \\
 	   10 & $\rho$  & LDA$\rightarrow$PBE  & 0.1 (0.2) & 0.2 (0.2) & 0.3 (0.6) & 0.2 (0.3) & 0.8 (1.4) \\ \hline
 		1 & $\rho$  & LDA$\rightarrow$MGGA  & 0.1 (0.1) & 0.2 (0.3) & 0.1 (0.2) & 0.1 (0.2) & 0.3 (0.3) \\
 		2 & $\rho$  & LDA$\rightarrow$MGGA  & 0.1 (0.2) & 0.3 (0.3) & 0.1 (0.2) & 0.1 (0.2) & 0.3 (0.5) \\
 		3 & $\rho$  & LDA$\rightarrow$MGGA  & 0.1 (0.2) & 0.3 (0.4) & 0.2 (0.3) & 0.2 (0.2) & 0.5 (0.7) \\
 	   10 & $\rho$  & LDA$\rightarrow$MGGA  & 0.1 (0.2) & 0.2 (0.3) & 0.4 (0.7) & 0.2 (0.3) & 0.9 (1.5) \\
	   
	\end{tabular}
\caption{\label{main_results_2_table} Mean absolute errors (in mHa per electron) and root mean squared errors (in parenthesis) for models trained in this report for the SHO potentials. The abbreviations $\rho$, and $V_{\text{ext}}$ are charge density, and potential respectively. The arrows (i.e. LDA$\rightarrow$PBE) indicate that the charge density used as input to the DNN was calculated using the LDA exchange-correlation functional, but the labels (energies) were calculated using another exchange-correlation functional.}
\end{table*}

\clearpage

\section{Mean absolute errors with the RND external potentials}
\label{mae_rnd}
\begin{table*}[ht]
	\begin{tabular}{c|c|c|c|c|c|c|c}
	$N_{\text{electrons}}$ & Input & Functional & $E_{\text{correlation}}$ &  $E_{\text{exchange}}$ & $E_{\text{external}}$ & $E_{\text{kinetic}}$ & $E_{\text{total}}$  \\ \hline
		1 & $\rho$  & LDA                   & 0.9 (1.5) & 1.4 (2.3) & 14.5 (31.7) & 2.5 (3.7) & 14.5 (31.1)  \\
		2 & $\rho$  & LDA                   & 1.1 (2.0) & 1.8 (3.1) & 9.6 (19.4) & 2.2 (3.4) & 9.9 (25.4)  \\
		3 & $\rho$  & LDA                   & 2.0 (3.5) & 3.1 (5.4) & 35.1 (57.0) & 5.6 (8.6) & 36.1 (58.4)  \\
	   10 & $\rho$  & LDA                  & 2.0 (3.8) & 3.0 (5.9) & 73.4 (117.1) & 6.7 (10.2) & 74.4 (119.4)  \\\hline
		1 & $\rho$  & PBE                  & 1.9 (2.5) & 3.1 (4.0) & 15.3 (32.2) & 3.8 (4.9) & 15.1 (31.2)  \\
		2 & $\rho$  & PBE                  & 1.3 (2.1) & 1.9 (3.1) & 9.5 (19.3) & 2.2 (3.1) & 10.1 (21.7)  \\
		3 & $\rho$  & PBE                  & 2.0 (3.3) & 2.9 (4.9) & 34.4 (55.6) & 5.2 (7.7) & 35.6 (56.8)  \\
	   10 & $\rho$  & PBE                  & 1.9 (3.6) & 2.8 (5.5) & 73.7 (115.0) & 7.6 (11.4) & 75.0 (117.1)  \\ \hline
		1 & $\rho$  & MGGA                 & 1.0 (1.7) & 2.1 (3.4) & 13.7 (30.7) & 2.5 (3.8) & 13.9 (31.0) \\
		2 & $\rho$  & MGGA                 & 1.3 (2.2) & 2.6 (4.6) & 10.0 (17.8) & 2.4 (3.9) & 10.2 (17.8)  \\
		3 & $\rho$  & MGGA                 & 1.8 (2.8) & 3.6 (5.7) & 34.1 (55.7) & 4.7 (7.1) & 36.5 (62.2)  \\
	   10 & $\rho$  & MGGA                 & 2.2 (4.1) & 4.4 (8.4) & 77.2 (122.1) & 7.3 (11.3) & 78.5 (126.1)  \\ \hline
		1 & $V_{\text{ext}}$ & LDA                  & 5.3 (9.0) & 8.6 (14.4) & 17.8 (28.7) & 11.5 (19.2) & 22.3 (36.8) \\
		2 & $V_{\text{ext}}$ & LDA                  & 5.3 (8.7) & 8.5 (13.6) & 18.5 (31.8) & 8.8 (13.5) & 21.2 (33.4) \\
		3 & $V_{\text{ext}}$ & LDA                  & 10.6 (15.0) & 16.8 (23.7) & 46.2 (70.2) & 15.8 (23.6) & 43.2 (66.7) \\
	   10 & $V_{\text{ext}}$ & LDA                  & 6.6 (9.9) & 10.3 (15.4) & 50.3 (86.1) & 10.5 (16.7) & 40.9 (73.2)  \\ \hline
		1 & $V_{\text{ext}}$ & PBE                  & 6.0 (9.9) & 9.3 (15.2) & 17.2 (27.3) & 12.1 (19.3) & 22.5 (35.6) \\
		2 & $V_{\text{ext}}$ & PBE                  & 6.0 (9.7) & 9.0 (14.7) & 19.5 (32.8) & 10.0 (15.4) & 21.3 (34.0) \\
		3 & $V_{\text{ext}}$ & PBE                  & 10.9 (15.3) & 16.8 (23.8) & 46.1 (69.5) & 16.3 (24.2) & 43.6 (65.4) \\
	   10 & $V_{\text{ext}}$ & PBE                  & 7.1 (10.6) & 10.8 (16.1) & 49.3 (85.4) & 10.4 (16.8) & 39.4 (71.5)  \\ \hline
		1 & $V_{\text{ext}}$ & MGGA                 & 5.7 (9.8) & 12.0 (20.4) & 16.1 (25.3) & 10.9 (17.9) & 21.7 (34.8)  \\
		2 & $V_{\text{ext}}$ & MGGA                 & 5.4 (8.6) & 11.0 (17.8) & 14.8 (23.2) & 8.1 (12.4) & 19.3 (30.0)  \\
		3 & $V_{\text{ext}}$ & MGGA                 & 11.1 (15.5) & 22.9 (32.3) & 44.5 (67.9) & 16.7 (25.0) & 43.0 (65.7)  \\
	   10 & $V_{\text{ext}}$ & MGGA                 & 7.3 (10.8) & 15.0 (22.2) & 50.8 (88.0) & 10.8 (16.9) & 41.4 (76.0)  \\ \hline
 		1 & $\rho$  & LDA$\rightarrow$PBE  & 1.2 (2.0) & 1.7 (3.0) & 14.5 (31.1) & 2.4 (3.6) & 14.6 (30.5)  \\
 		2 & $\rho$  & LDA$\rightarrow$PBE  & 1.2 (4.3) & 1.8 (6.4) & 8.6 (19.3) & 1.9 (4.6) & 9.1 (23.9)  \\
 		3 & $\rho$  & LDA$\rightarrow$PBE  & 1.9 (3.4) & 2.8 (5.0) & 34.8 (56.2) & 5.1 (7.7) & 35.6 (56.9)  \\
 	   10 & $\rho$  & LDA$\rightarrow$PBE  & 2.2 (4.0) & 3.2 (6.1) & 75.5 (118.4) & 7.6 (11.7) & 76.7 (120.9)  \\ \hline
 		1 & $\rho$  & LDA$\rightarrow$MGGA & 1.4 (2.8) & 2.9 (5.8) & 14.2 (30.9) & 2.6 (4.5) & 14.7 (31.6) \\
 		2 & $\rho$  & LDA$\rightarrow$MGGA & 1.5 (3.7) & 3.0 (7.5) & 8.0 (15.2) & 2.0 (4.3) & 8.7 (19.6)  \\
 		3 & $\rho$  & LDA$\rightarrow$MGGA & 2.9 (5.4) & 6.0 (11.3) & 36.3 (58.2) & 5.8 (10.0) & 37.7 (59.8)  \\
 	   10 & $\rho$  & LDA$\rightarrow$MGGA & 2.6 (4.9) & 5.3 (9.9) & 73.6 (115.6) & 7.5 (11.7) & 74.4 (117.4)  \\ \hline
	   10 & $V_{\text{ext}}\rightarrow\rho$ & LDA  & 6.4 (11.0) & 9.9 (17.1) & 93.0 (151.4) & 12.7 (21.5) & 98.9 (167.5) \\
	   10 & $V_{\text{ext}}\rightarrow\rho$ (DDE) & LDA  & 6.4 (10.7) & 10.0 (16.7) & 99.4 (154.4) & 13.4 (22.2) & 108.2 (175.5) \\
	   10 & Predicted $\rho$ & LDA & 5.8 (10.1) & 9.0 (15.7) & 91.4 (143.2) & 11.2 (19.1) & 98.5 (158.7) \\
	   10 & $V_{\text{ext}}\rightarrow\rho$ & PBE & 7.2 (11.7) & 10.9 (17.7) & 100.4 (164.8) & 14.5 (24.6) & 108.1 (185.8)  \\
	   10 & $V_{\text{ext}}\rightarrow\rho$ (DDE) & PBE  & 7.2 (11.5) & 10.9 (17.7) & 107.0 (168.1) & 15.1 (24.9) & 117.6 (193.7)  \\
	   10 & Predicted $\rho$ & PBE & 6.8 (11.1) & 10.2 (16.8) & 98.2 (151.5) & 12.4 (21.4) & 106.7 (170.3) \\
	   10 & $V_{\text{ext}}\rightarrow\rho$ & MGGA & 7.5 (12.3) & 15.4 (25.1) & 94.6 (161.9) & 13.0 (22.8) & 103.3 (180.5)  \\
	   10 & $V_{\text{ext}}\rightarrow\rho$ (DDE) & MGGA & 7.4 (12.0) & 15.2 (24.6) & 101.1 (162.4) & 13.7 (23.3) & 111.8 (184.0)\\
	   10 & Predicted $\rho$ & MGGA & 7.1 (12.1) & 14.7 (24.8) & 94.9 (150.2) & 12.7 (21.5) & 102.5 (167.3)
	   
	\end{tabular}
\caption{\label{main_results_1_table} Mean absolute errors (in mHa per electron) and root mean squared errors (in parenthesis) for models trained in this report for the RND potentials. The abbreviations $\rho$, $V_{\text{ext}}$, and Predicted $\rho$ are charge density, potential, and predicted charge density respectively. The arrows (i.e. LDA$\rightarrow$PBE) indicate that the charge density used as input to the DNN was calculated using the LDA exchange-correlation functional, but the labels (energies) were calculated using another exchange-correlation functional. The acronym DDE stands for density driven error, as defined by Brockherde \emph{et al.} \cite{BVLT17}. The models labelled by $V_{\text{ext}}\rightarrow\rho$ directly map the external potentials to charge densities. The models labelled by Predicted $\rho$ map predicted charge densities to true energies.}
\end{table*}

\clearpage

\section{A note on the density driven errors}
\label{MADDE}
In Table \ref{main_results_1_table}, we report the mean absolute density driven error rather than the density driven error that is reported in \cite{BVLT17}. When evaluating a machine learning model, it is common to report absolute errors to avoid reporting an average error that would have error cancellation. Consider the total energy expression
\begin{equation}
  E[\rho] = F[\rho] + \int d{\bf r}~V({\bf r})\rho({\bf r})
\end{equation}
where $F$ is the universal functional defined in \cite{kohn1965self} and $V$ is the external potential. The value $E$ is the \emph{true} energy given a \emph{true} charge density $\rho$. The total error reported in \cite{BVLT17} is defined to be 
\begin{equation}
  \Delta E = \tilde{E}[\tilde{\rho}] - E[\rho] = \Delta E_F + \Delta E_D
\end{equation}
where 
\begin{equation}
\Delta E_F = \tilde {F}[\rho] - F[\rho]
\end{equation}
is the functional driven error and 
\begin{equation}
  \Delta E_D = \tilde {E}[\tilde{\rho}] - \tilde{E}[\rho]
\end{equation}
is the density driven error. The variables $\tilde{\rho}$, $\tilde{E}$, and $\tilde{F}$ represent a predicted charge density, an approximation to the total energy functional, and an approximation to true universal functional, respectively. If we consider the absolute value of the total error 
\begin{eqnarray}
  |\Delta E| = |\Delta E_F + \Delta E_D|\neq |\Delta E_F| + |\Delta E_D|
\end{eqnarray}
then we can see that there must be error cancellation between the terms $\Delta E_F + \Delta E_D$ in order for $|\Delta E| < |\Delta E_D|$. This is what we find in Table \ref{main_results_1_table}.
\end{document}